# An Overview of the Commercial Cloud Monitoring Tools: Research Dimensions, Design Issues, and State-of-the-Art

Khalid Alhamazani, University of New South Wales, Australia
Rajiv Ranjan, CSIRO, Australia
Karan Mitra, CSIRO, Australia
Fethi Rabhi, University of New South Wales, Australia
Samee Ullah Khan, North Dakota State University, USA
Adnene Guabtni, NICTA, Australia
Vasudha Bhatnagar, University of Delhi, India

**Abstract.** Cloud monitoring activity involves dynamically tracking the Quality of Service (QoS) parameters related to virtualized resources (e.g., VM, storage, network, appliances, etc.), the physical resources they share, the applications running on them and data hosted on them. Applications and resources configuration in cloud computing environment is quite challenging considering a large number of heterogeneous cloud resources. Further, considering the fact that at each point of time, there will be a different and specific cloud service which may be massively required. Hence, cloud monitoring tools can assist a cloud providers or application developers in: (i) keeping their resources and applications operating at peak efficiency; (ii) detecting variations in resource and application performance; (iii) accounting the Service Level Agreement (SLA) violations of certain QoS parameters; and (iv) tracking the leave and join operations of cloud resources due to failures and other dynamic configuration changes.

In this paper, we identify and discuss the major research dimensions and design issues related to engineering cloud monitoring tools. We further discuss how aforementioned research dimensions and design issues are handled by current academic research as well as by commercial monitoring tools.

## 1. Introduction

According to National Institute of Standards and Technology NIST[1], cloud computing is a "*Model for enabling convenient, on-demand network access to a shared pool of configurable computing resources (network, servers, storage, applications, services) that can be rapidly provisioned and released with minimal management effort or service provider interaction*" [27]. *Service models*, *hosting*, *deployment models*, and *roles* are some of the important concepts related to cloud technology, defined by NIST and the cloud community. These are essential characteristics, that have been elaborated in [27][28][29][30][31], and [32]. Commercial cloud providers including Amazon Web Services (AWS), Microsoft Azure, Salesforce.com, Google App Engine and others offer the cloud consumers options to deploy their applications over a network of infinite resource pool with practically no capital investment and with modest operating cost proportional to the actual use. For example, Amazon EC2 cloud runs around half million physical hosts, each of them host multiple virtual machines that can be dynamically invoked or removed [12].

Several papers in literature discuss, explore and propose surveys of cloud monitoring in different aspects [13] [15] [16] [17] [27][28][29][30][31][32]. To the best of our knowledge, no specific survey considers monitoring applications at different cloud layers (Infrastructure as a service, platform as a service and software as a service). Further, none of the papers has focused on predictive cloud monitoring. In addition to that, none of the paper discusses the possibility of utilizing machine learning techniques with monitored data. In addition to the above factors, one arising aspect with the cloud computing is managing huge volume of data (Big Data). In the present environment, the term "Big Data" is described as a phenomenon that refers to the practice of collection and processing of very large datasets and the associated systems and algorithms used to analyze enormous those data sets [53]. Three well recognized characteristics of Big data are Variety, Volume and Velocity (3 V's) of data generation [46][47]. The steady growth of social media and mobile devices has led to an increase in the sources of outbound traffic, initiating "data tsunami phenomenon". This poses significant challenges in cloud computing.

---

[1] http://www.nist.gov/itl/cloud/

In [50][51][52], studies show that as more people join the social media sites hosted on clouds, analysis of the data becomes more difficult and almost impossible to be analyzed. Another aspect of big data events can occur by the cloud infrastructure itself [49]. Other study [48] shows that VMs migrating, copy and saving current state can affect the performance of data transfer within the cloud. Moreover, different types of data originating from mobile devices makes understanding of composite data a challenging problem due to multi-modality, huge volume, dynamic nature, multiple sources, and unpredictable quality. Continuously monitoring of multi-modal data streams collected from heterogeneous sources require monitoring tools can cope up with managing big data floods.

In this paper, we identify and discuss three challenges of cloud monitoring: (1) How would application monitoring be a layer specific i.e., how cloud consumer can stipulate at what cloud layer his/her running application should be monitored. (2) How cloud consumer can express what information he/she is interesting in to gain knowledge while his/her application is being monitored. (3) What is the possibility for the cloud consumer to have a predictive status about his/her running application in future, which leads to gain more awareness about the application future behavior.

## 1.1. **Our Contributions**

The concrete contributions made by this paper are: (a) *advancing the fundamental understanding of cloud resource and application provisioning and monitoring concepts*; (b) *identification of the main research dimensions and programming issues based on cloud resource and application types*; and (c) *presents future research directions for novel cloud monitoring techniques.*

The remainder of the paper is organized as follows: Discussion on key cloud resource provisioning is presented in Section 2. Section 3, discusses the cloud application life cycle and in detail discusses the components of cloud monitoring. Furthermore, details on major research dimensions and programming issues related to engineering cloud monitoring tools are presented in Section 4. Finally, mapping of research dimensions to existing cloud monitoring tools is discussed in Section 5. The paper ends with brief conclusion and overview of future work.

# 2. Cloud Resource Provisioning

Cloud resource provisioning is a complex task [8] and is referred to as the process of application deployment and management on cloud infrastructure [8]. Current cloud providers such as Amazon EC2, ElasticHosts, GoGrid, TerraMark, and Flexian, do not completely offer support for automatic deployment and configuration of software resources [14]. Therefore, several companies, e.g. RightScale and Scalr provide scalable managed services on top of cloud infrastructures, to cloud consumers [14]. Three main steps for cloud provisioning are [8][9]:

*Virtual Machine Provisioning* – where suitable VMs are instantiated to match the required hardware and configuration of an application. To illustrate, Bitnami[2] supports consumers to provision a Bitnami stack that consists of VM and appliances. On the other hand, Amazon EC2 consumers may firstly provision a VM on the cloud then choose the appliances to provision on that VM.

*Resource Provisioning* – it is the process of mapping and scheduling the instantiated VMs onto the cloud physical servers. This is handled by cloud-based hypervisors. For example, public clouds expose APIs to start/stop a resource but not to control which physical server within that region/datacenter will host the VM. Figure 1, illustrates the steps where a cloud consumer attempts to provision cloud resources on the cloud Amazon EC2 platform. In step 1, from the VM repository, a consumer views the available VMs provided by the cloud platform and selects the preferable VM instance type. In step 2, the consumer sets up his/her preferences/configurations on this VM. In steps 3 and 4, the user deploys this VM on the cloud platform successfully. Subsequently, in steps 5 and 6, the consumer retrieves back a list of available applications from the applications repository. In step 7, the consumer simply opts for his/her desired applications that he/she would like to provision. Finally, in step 8, the cloud consumer deploys the applications and the VM on the cloud platform.

[2] http://bitnami.org/faq/cloud_amazon_ec2

*Application Provisioning* – is the process of application deployment on VMs on the cloud infrastructure. For example, deploying a Tomcat server as an application on a VM hosted on the Amazon EC2 cloud. Applications provisioning can be done in two ways. The first method consists of deploying the applications together while hosting a VM. In the second method, the consumer may want to first deploy the VM, and then as a separate step, he/she may deploy the applications.

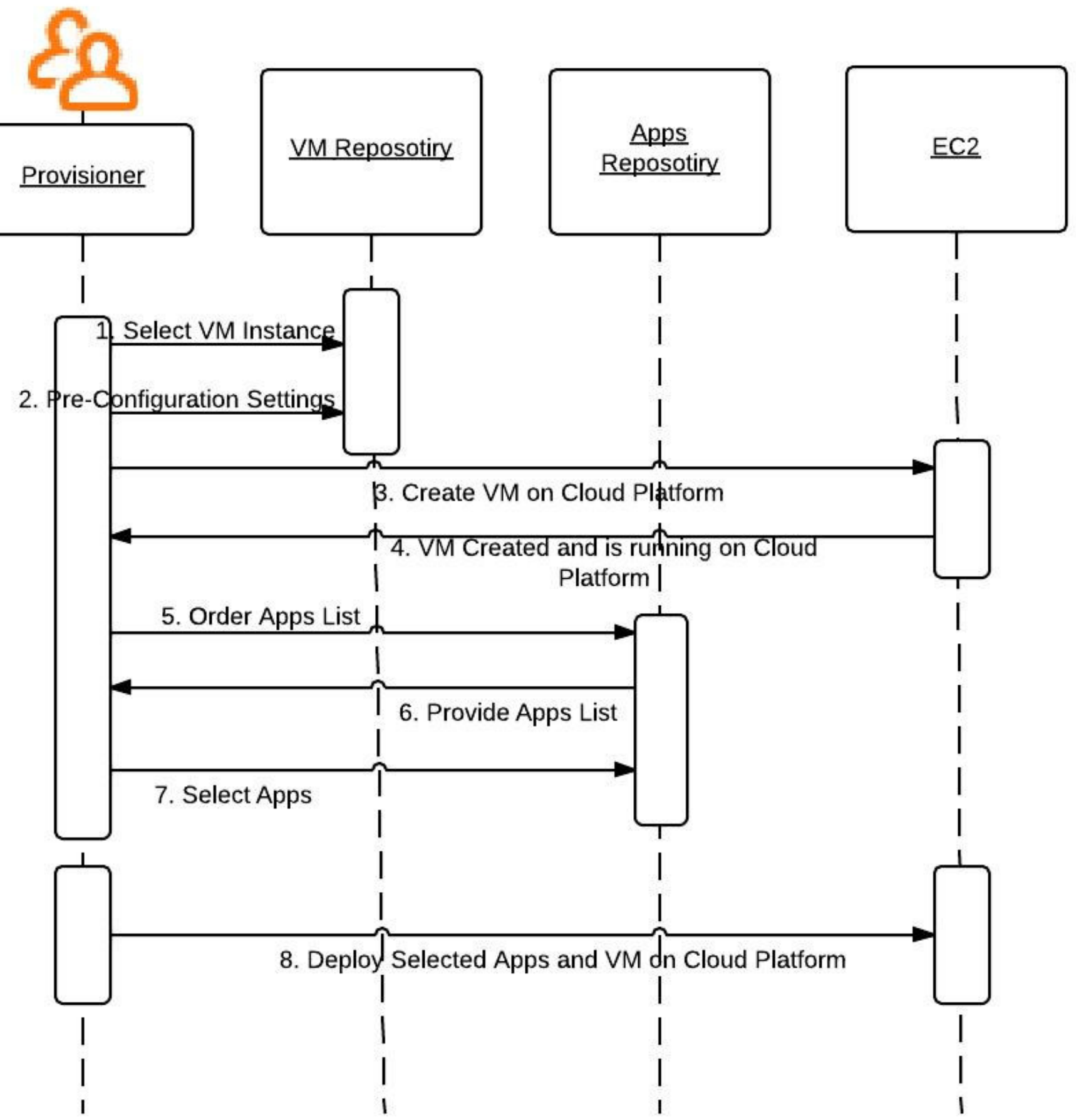

**Fig. 1**: Provisioning and Deployment Sequence Diagram

After the provisioning stage, cloud workflow instance might be composed of multiple cloud services, and in some cases from a number of different service providers. Therefore, monitoring the quality of cloud instance becomes a much more complex [10]. Further, at the instance run time, QoS of the running instance needs to be consistently monitored to guarantee avoiding and handling violations. Monitoring is the process of observing and tracking applications/resources at run time. It is the basis of control operations and corrective actions for running systems on clouds. Despite the existence of many commercial monitoring tools available in the market, SLAs between providers and consumers still pose a major issue in clouds.

In some way, cloud monitoring, service level agreements (SLA) and dynamic configuration are correlated in the sense that one has an impact on another. In other words, enhancing monitoring functionalities will in turn assist meeting SLAs as well as improving dynamic configuration operations at run time. Moreover, SLA has to be met by the cloud providers in order to reach the required reliability level required by consumers. Also, auto-scaling and dynamic configurations are required with cloud technology. This all-together leads us conclude that the monitoring process is the key element that has to be further studied and enhanced to upgrade the quality level of the aforementioned factors.

## 3. Cloud Monitoring

Under this section, we present the basic components, phases and layers of application architecture on clouds. Also, this section will present the state of the art of cloud monitoring as well as how it is conceptually correlated to QoS and SLA.

### 3.1. Application Life Cycle

The application architecture determines how, when, and which provisioning operations should be affected on cloud resources. The high level application (e.g., multimedia applications) architecture is multi-layered [54]. These layers may consist of clients/application consumers, load balancers, web servers, streaming servers, application servers, and a database system. Notably, each layer may instantiate multiple software resources as and when required and such multiple instantiations can be allocated to one or more hardware resources. Technically, across those aforementioned system layers, a number of provisioning operations take place at design time as well as run-time. These operations should ensure the SLA compliance by achieving the quality of service (QoS) targets.

*Resource Selection* – It is the process where the system developer selects software (web server, multimedia server, database server, etc.) and hardware resources (CPU, storage, and network). This process encapsulates the allocation of hardware resources to those selected software resources.

*Resource Deployment* – During this process, system administrator instantiates the selected software resources on the hardware resources, as well as configuring these resources for successful communication and inter-operation with the other software resources already running in the system.

*Resource Monitoring* - In order to ensure that the deployed software and hardware resources run at the required level to satisfy the SLA, a continuous resource monitoring process is desirable. This process involves detecting and gathering information about the running resources. In case of the detection of any abnormal system behavior, the system orchestrator is notified for policy-based corrective actions to be undertaken as a system remedy.

*Resource Control* – Is the process to satisfy system technical requirements by ensuring to meet the QoS stated in the SLA. It is responsible for handling system uncertainties at run time e.g. upgrade or downgrade a resource type or functionality.

### 3.2. Cloud Monitoring

In clouds, monitoring is essential for the health of the system and is important for both providers and consumers [13] [15] [16] [17]. Primarily, monitoring is a key tool for i) managing software and hardware resources; and ii) providing continuous information for those resources as well as for consumers' hosted applications on the cloud. Cloud activities like resource planning, resource management, data center management, SLA management, billing, troubleshooting, performance management, and security management essentially need monitoring to effective and smooth operations of the system [18]. Consequently, there is a strong need for monitoring looking at the elastic nature of cloud computing [1].

In cloud computing, monitoring can be of two types: high-level and low-level. High-level monitoring is related to the virtual platform status [20]. The low-level monitoring is related to information collected for the status of the physical infrastructure [20] [21]. Cloud monitoring system is a self-adjusting and typically multi-threaded system that is able to support monitoring functionalities [11]. It comprehensively monitors pre-identified instances/resources on the cloud for abnormalities. On detecting an abnormal behavior, the monitoring system attempts to auto-repair this instance/resource if the corresponding monitor has a tagged auto-heal action [11]. In case of auto-repair failure or an absence of an auto-heal action, a support team is notified. Technically, notifications can be sent by different means such as email, or SMS [11].

### 3.2.1. **Monitoring, QoS, and SLA**

As mentioned earlier, cloud monitoring is needed for continuous measurements to assess resources or applications on cloud platform in terms of performance, reliability, power usage, ability to meet SLA, security, etc. [19]. Fundamentally, monitoring tests can be computation based and/or network based. Computation based tests are concerned about the status of the real or virtualized platforms running cloud applications. Data metrics considered in such test include CPU speed, CPU utilization, disk throughput, VM acquisition/release time and system up-time. Network based tests focus on network layer data related metrics like jitter, round-trip time RTT, packets loss, traffic volume etc. [20][21][23][24].

At run-time, a set of operations takes place in order to meet the QoS parameters specified in SLA document that guarantees the required performance objectives of the cloud consumers. The availability, load, and throughput of hardware resources can vary in unpredictable ways, so ensuring that applications achieve QoS targets is not trivial. Being aware of the system's current software and hardware service status is imperative for handling such uncertainties to ensure the fulfillment of QoS targets [13]. In addition, detecting exceptions and malfunctions while deploying software services on hardware resources is essential e.g., showing QoS delivered by each application component (software service such as web server or database server) hosted on each hardware resource. Uncertainties can be tackled through the development of efficient, scalable, interoperable monitoring tools with easy-to-use interfaces.

### 3.2.2. **Monitoring across different applications and Layers**

As mentioned previously, application components (e.g., streaming server, web server, indexing server, compute service, storage service, and network) are distributed across cloud layers including PaaS and IaaS. Thus, in order to guarantee the achievement of QoS targets for the application as a whole, monitoring QoS parameters should be performed across all the layers of cloud stack including Platform-as-a-Service (PaaS) (e.g., web server, streaming server, indexing server, etc.) and Infrastructure-as-a-Service (IaaS) (e.g., compute services, storage services, and network). Figure 2 illustrates how different components in a cloud platform are distributed across the cloud platform layers. Table 1 shows the QoS parameters that a monitoring system should be considering at each cloud layer.

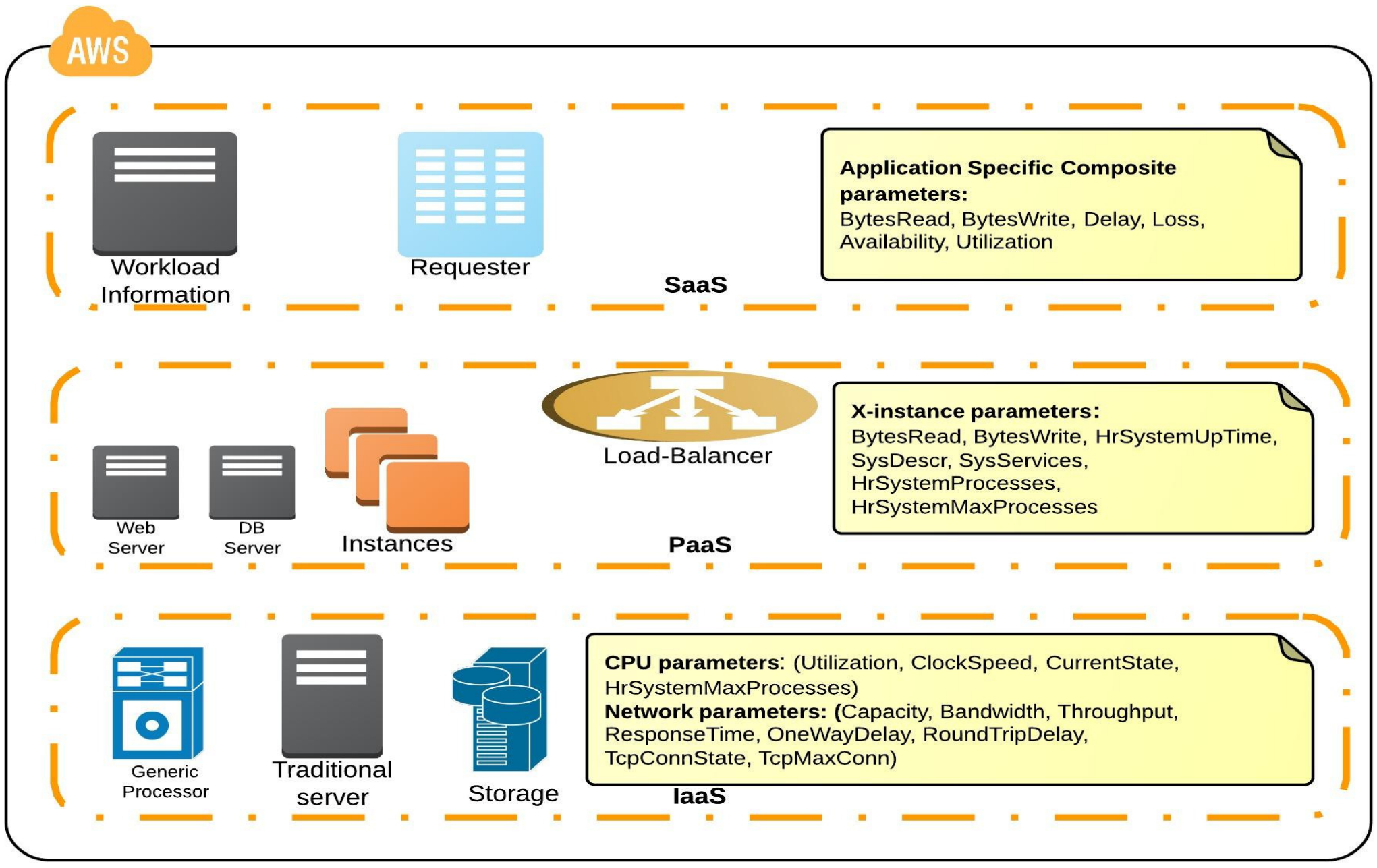

**Fig. 2**: **Components across cloud platform layers.**

**Table 1: QoS parameters at each cloud platform layer**.

| Cloud Layer | Layer Components | Targeted QoS Parameters |
|---|---|---|
| **SaaS** | Appliances x,y,z, etc. | BytesRead, BytesWrite, Delay, Loss, Availability, Utilization. |
| **PaaS** | Web Server, Streaming Server, Index Server, Apps Server, etc. | BytesRead, BytesWrite, SysUpTime, SysDesc, HrSystemMaxProcesses, HrSystemProcesses, SysServices. |
| **IaaS** | Compute Service, Storage Service, Network, etc. | CPU Parameters: Utilization, ClockSpeed, CurrentState<br>Network Parameters: Capacity, Bandwidth, Throughput, ResponseTime, OneWayDelay, RoundTripDelay, TcpConnState, TcpMaxConn |

Typically, QoS targets vary across application types. For example, QoS targets for eResearch applications are different from static, single-tier web applications (e.g., web site serving static contents) or multi-tier applications (e.g., on demand audio/video streaming). Based on application types, there is always a need to negotiate different SLAs. Hence, SLA document includes conditions and constraints that match the nature of QoS requirements with each application type. For example, a genome analysis experiment on cloud services will only care of data transfer (upload and download) network latency and processing latency. On the other hand, for multimedia applications, the quality of the transferred data over network is more important. Hence, other parameters gain priority in this case. Failing to track QoS parameters will eventually lead to SLA violations. Consequently, monitoring is fundamental and responsible for SLAs compliance certification [26]. Moreover, monitoring definitively will enable cloud providers to frame more realistic and dynamic SLAs models by getting advantage of the knowledge of consumer-perceived performance [25].

## 4. Evaluation Dimensions

Under this section, we present the basic components that can be considered as evaluation dimensions in order to evaluate a monitoring tool in cloud computing.

### 4.1. Monitoring Architectures

In cloud monitoring, the network and system related information is collected by the systems. For example, CPU utilization, network delay and packet losses. This information is then used by the applications to determine actions such as data migration to the server closest to the user to ensure that SLA requirements are met. Typically, network monitoring can be performed on centralized and de-centralized network architectures.

#### 4.1.1. **Centralized**

In centralized architecture shown in figure 3, the PaaS and IaaS resources send QoS status update queries to the centralized monitoring server. In this scheme, the monitoring techniques continuously pull the information from the components via periodic probing messages. In [11], the authors show that a centralized cloud monitoring architecture allows better management for cloud applications. Nevertheless, centralized approach has several design issues, including:

- Prone to a single point of failure;
- Lack of scalability;
- High network communication cost at links leading to the information server (i.e., network bottleneck, congestion); and
- Possible lack of the required computational power to serve a large number of monitoring requests.

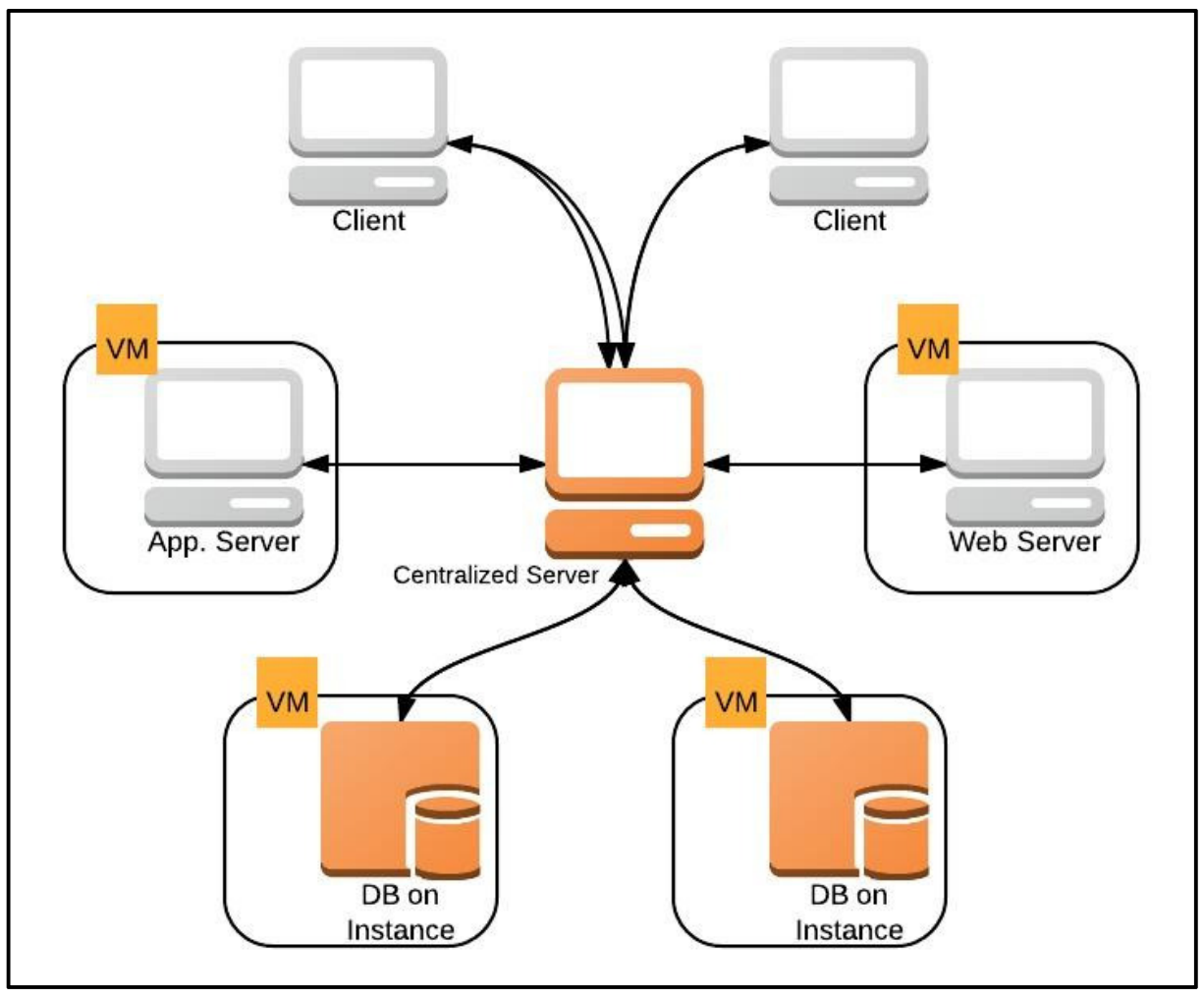

**Fig. 3: Centralized Monitoring Architecture**

### 4.1.2. Decentralized

Recently, proposals for decentralized cloud monitoring tools have gained momentum. Figure 4 shows the broad schematic design of decentralized cloud monitoring system. The decentralization of monitoring tools can overcome the issues related to current centralized systems. A monitoring tool configuration is considered as decentralized if none of the components in the system is more important than others. In case one of the components fails, it does not influence the operations of any other component in the system.

*Structured peer-to-peer* - Looking forward for having a network layout where a central authority is defused has to the development of the structured peer-to-peer networks. In such a network overlay, central point of failure is eliminated. Napster is a popular structured peer-to-peer system [4] [5] [6].

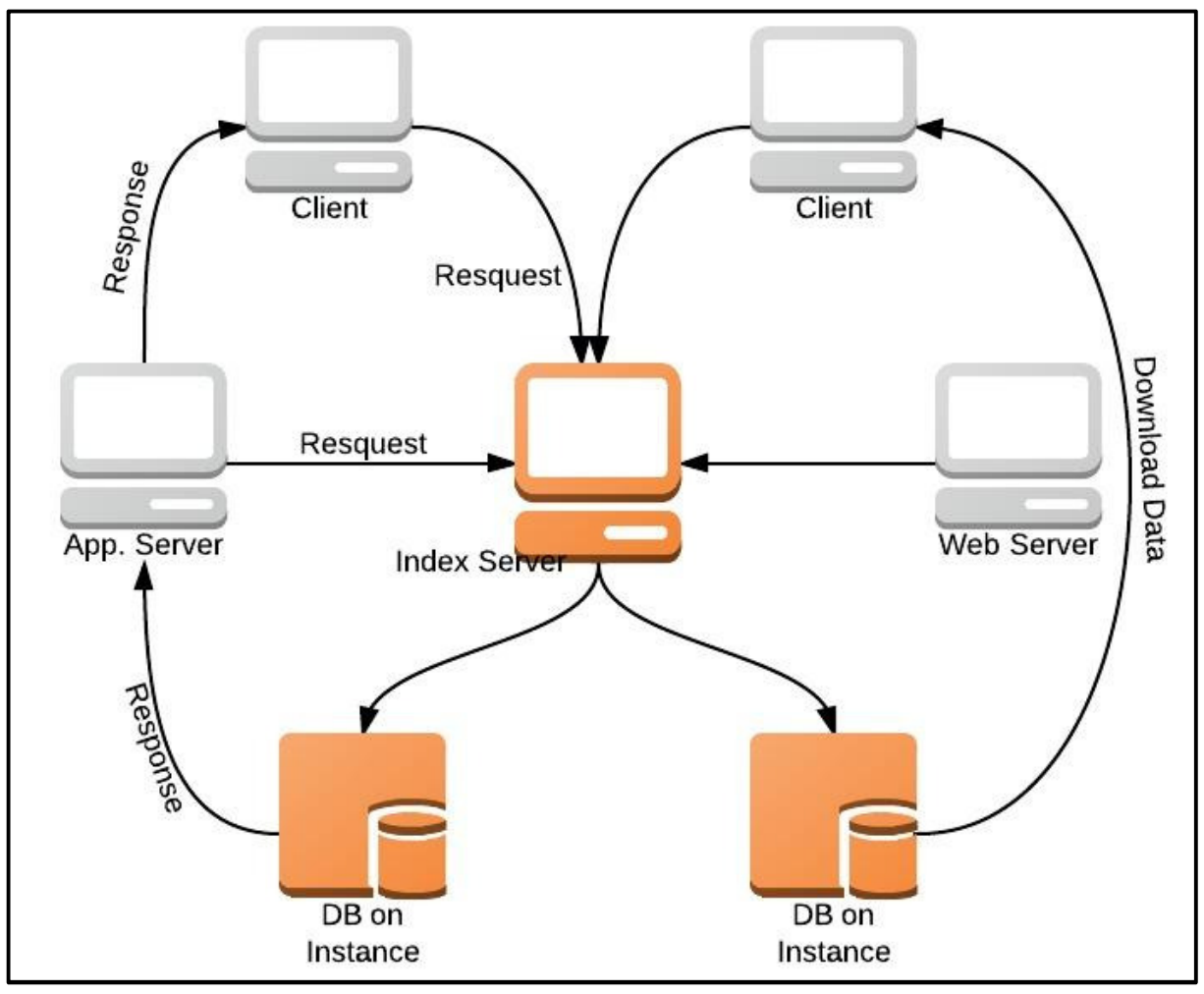

**Fig. 4: Decentralized Monitoring Architecture**

*Unstructured peer-to-peer* - Unstructured peer-to-peer networks overlay is meant to be a distributed overlay but the difference is that the search directory is not centralized unlike structured peer-to-peer networks overlay. This leads to absolute single point failure in such network overlay. Gnutella is one of the well-known unstructured peer-to-peer systems [6] [7].

*Hybrid peer-to-peer* - Is a combination of structured and unstructured peer-to-peer networks systems. Super peers can act as local search hubs in small portions of the network where the general scope of the network behaves as unstructured peer-to-peer system. Kazaa is a hybrid of centralized Napster and decentralized Gnutella network systems.

## 4.2. **Interoperability**

The interoperability perspective in technology focuses on the system's technical capabilities to interface between organizations and systems. It also focuses on the resulting mission of compatibility or incompatibility between systems and data collation partners. Modern business applications developed on cloud are often complicated and require interoperability. For example, an application owner can deploy a web server on Amazon Cloud while the database server may be hosted in Azure Cloud. Unless data and applications are not integrated across clouds properly, the results and benefits of cloud adoption cannot be achieved. Interoperability is also necessary to avoid cloud provider lock-in.

This dimension refers to the ability of a cloud monitoring framework to monitor applications and its components that may be deployed across multiple cloud providers. While it is not difficult to implement a cloud-specific monitoring framework, to design generic monitoring framework that can work with multiple clouds remains a challenging problem. Next, we classify the interoperability (figure 5) of monitoring frameworks into the following categories:

*Cloud Dependent* – Currently many public cloud providers provide ability to their consumers to monitor their applications using available monitoring tools for CPU, storage and network. Often these tools are tightly integrated with their other existing cloud. For example, CloudWatch, offered by Amazon is a monitoring tool that enables consumers to manage and monitors their applications residing on AWS EC2 (CPU) services. But, this monitoring tool does not have the ability to monitor an application component that may reside on other cloud provider's infrastructure such as GoGrid and Azure. Table 2 illustrates some examples of cloud monitoring tools that are specific to a cloud provider as well as it shows those which are not, but are Cloud Agnostic.

**Table 2: Monitoring Tools and Interoperability**

| **Platform** | **Interoperability** Cloud-Agnostic (Multi-Clouds) |
|---|---|
| Monitis [33] | Yes |
| RevealCloud [36][37] | Yes |
| LogicMonitor [35] | Yes |
| Nimsoft [34] | Yes |
| Nagios [20][38] | Yes |
| SPAE [39][40] | Yes |
| CloudWatch [41] | No |
| OpenNebula [42] | No |
| CloudHarmony [43] | Yes |
| Azure FC [44][45] | No |

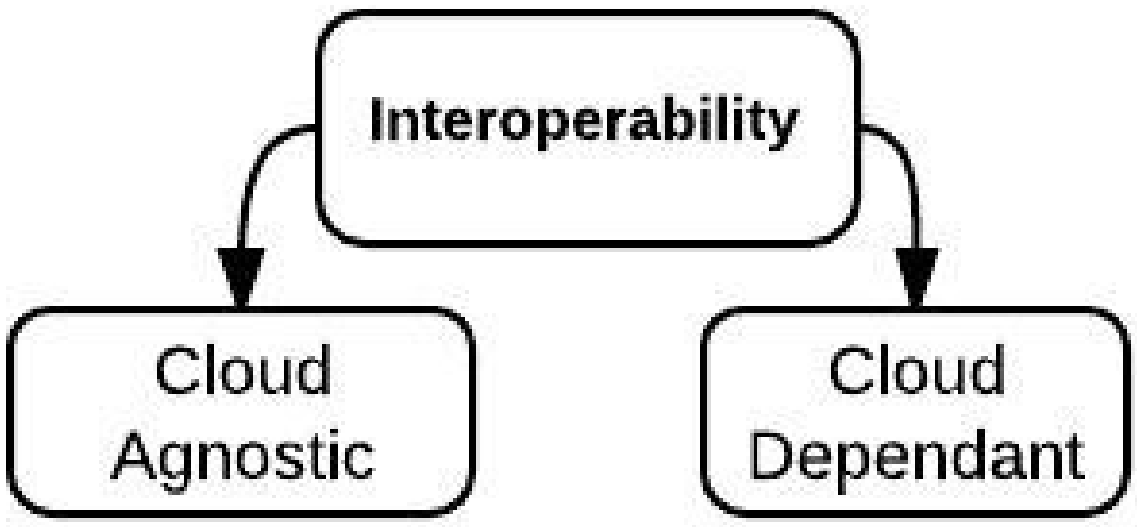

**Fig. 6: Interoperability Classification**

*Cloud Agnostic* – In contrast to single cloud monitoring, engineering cloud agnostic monitoring tools is challenging. This is primarily due to fact that there isn't a common unified application programming interface (API) for calling on cloud computing services' runtime QoS statistics. Though recent developments in cloud programming API including Simple Cloud, Delta Cloud, JCloud, and Dasein Cloud simplify interaction of services (CPU, storage, and network) that may belong to multiple clouds, they have limited or no ability to monitor their run-time QoS statistics and application behaviors. In this scenario, monitoring tools are expected to be able to retrieve QoS data of services and applications that may be part of multiple clouds. Cloud agnostic monitoring tools are also required if one wants to realize a hybrid cloud architecture involving services from private and public clouds. Monitis monitoring tool provides the ability of accessing different clouds e.g. Amazon EC2, Rackspace and GoGrid. It utilizes the concept of widgets where consumers can view more than one widget in a page. In Monitis, consumers need to provide the only cloud account credentials to access monitoring data of their different cloud applications. They can also specify which instance to monitor. Hence, a consumer can view two different cloud instances using two different widgets in one single page.

## 4.3. Quality of Service Matrix

It is non-trivial for application developers to understand what QoS parameters and targets he/she needs to specify and monitor across each layer of a cloud stack including PaaS (e.g., web server, streaming server, indexing server, etc.) and IaaS (e.g., compute services, storage services, and network). As shown in figure 6, this can be by one parameter or a group of parameters.

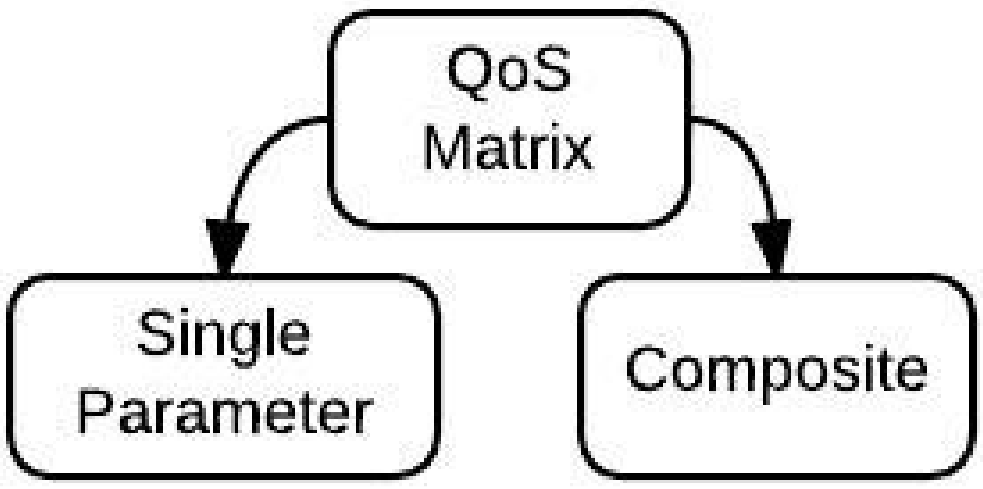

**Fig. 6: QoS matrix Classification**

### 4.3.1. Single Parameter

In this scenario, a single parameter refers to a specific system QoS target. In each system, there are major atomic/single values that have to be tracked closely and continuously. For example, CPU utilization is basically expressed by only one single parameter as well as in the SLA terms. Such values can affect the whole system and a violation to such value in SLA can lead to a serious system failure. Unlike other parameters' values that might be part of composite parameters where they might not present priority to the system administrator, single parameters gain in most cases high priority when monitoring SLA violations and QoS targets.

### 4.3.2. Composite Parameters

In a composite parameter scenario, a group of different parameters are taken into consideration. In the cloud, cloud software application is composed of many cloud software services. Thus, the performance quality can be determined by collective behaviors of those software services [10]. After observing multiple parameters for estimating a functionality of one or more concerned processes, one result could be obtained to evaluate the QoS. To illustrate, "loss" can be considered as a composite parameter of two single parameters "one way loss" and "round trip loss". Similarly, "delay" can be considered as composite parameters three single parameters "one way delay", "RTT delay", and "delay variance". Table 3 shows a list of some commercial tools for cloud monitoring and it illustrates which of them support or do not support monitoring multiple QoS parameters.

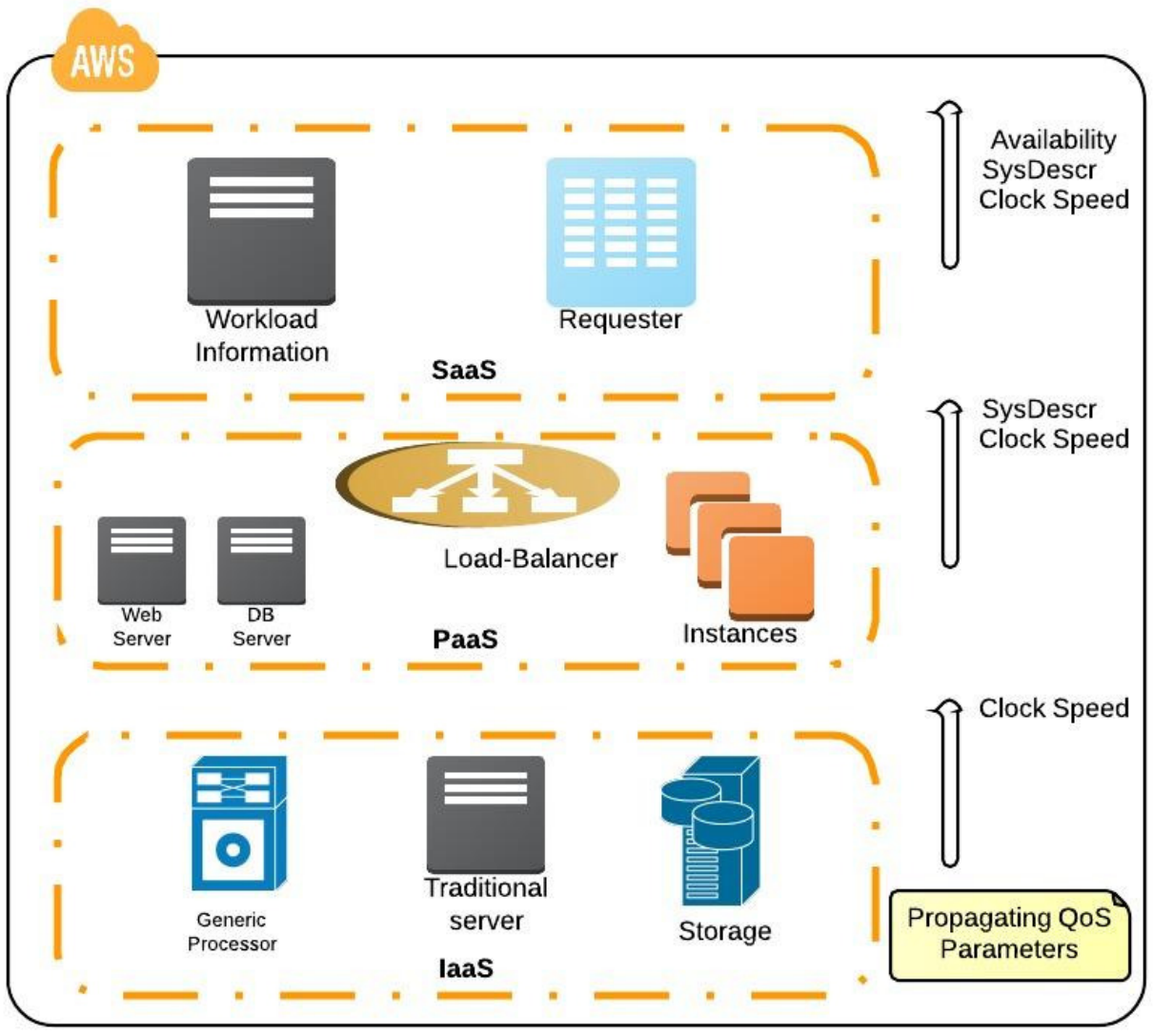

**Fig. 7: Components across Cloud layers and QoS Propagating**

## 4.4. **Visibility**

As shown in figure 7, application components (streaming server, web server, indexing server, compute service, storage service, and network) related to multimedia streaming application are distributed across cloud layers including PaaS and IaaS. Thus, in order to guarantee the achievement of QoS targets for the application as a whole, it is critical to monitor QoS parameters across multiple layers [3]. Hence, the challenge here is to develop monitoring tools that can capture and reason about the QoS parameters of application components across IaaS and PaaS layers. As demonstrated in figure 8, we categorize the visibility of commercial monitoring tools into following categories:

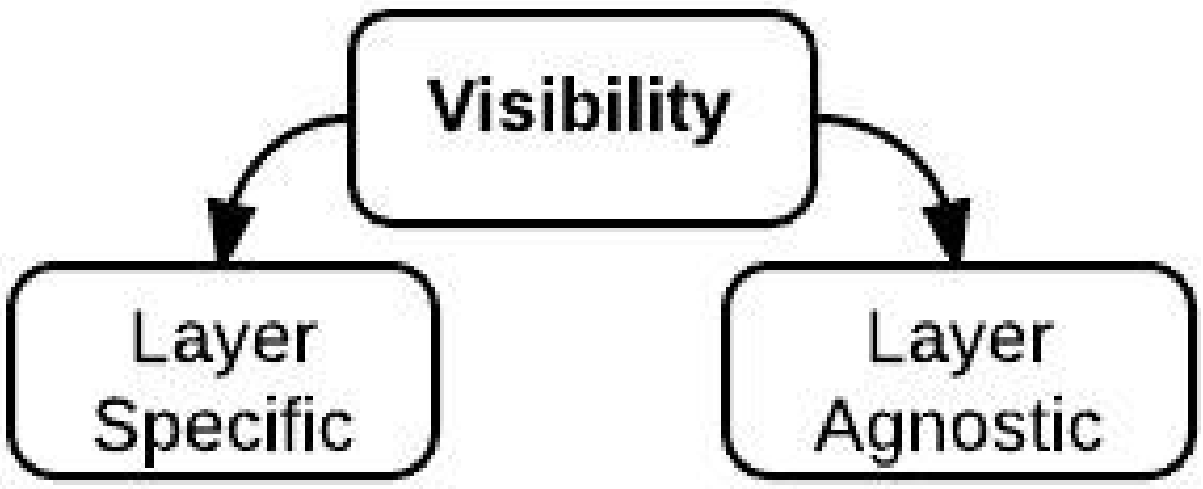

**Fig. 8: Visibility Categorization**

*Layer specific* – Cloud services are distributed among three basic layers, SaaS, PaaS, and IaaS. Monitoring tools originally are oriented to perform monitoring tasks over services only in one of the aforementioned layers. Most of present day commercial tools are designed to keep track of the performance of resources provisioned at the IaaS layer. For example, CloudWatch is not capable of monitoring information related to load, availability, and throughout of each core of CPU services and its effect on the QoS (e.g., latency, availability, etc.) delivered by the hosted PaaS services (e.g., J2EE application server). Hence, there exists a considerable research challenges in developing a monitoring tool that can monitor QoS statistics across multiple layers of the cloud stack.

*Layer Agnostic* – In contrast to the previous scenario, monitoring at multiple layers enables the consumers to gain access to applications' data across multiple layers. E.g., consumers can retrieve data at the same time from PaaS and IaaS for the same application (table 3). This type of cloud monitoring is essential in all cases but obviously it is more effective for consumers requiring a complete awareness about their running cloud applications.

**Table 3: Monitoring Tools and Layers' Visibility**

| Platform | Visibility<br>Multi-Layers<br>(Composite QoS Matrix) |
|---|---|
| Monitis [33] | Yes |
| RevealCloud [36][37] | Yes |
| LogicMonitor [35] | Yes |
| Nimsoft [34] | Yes |
| Nagios [20][38] | Yes |
| SPAE [39][40] | No |
| CloudWatch [41] | Yes |
| OpenNebula [42] | No |
| CloudHarmony [43] | No |
| Azure FC [44][45] | Yes |

## 4.5. **Programming Interfaces**

Programing interfaces allows the development of software systems to enable monitoring across different layers of the cloud stack. It involves several components such as APIs, widgets and the command line to enable a consumer to monitor several components of the complex cloud systems in a unified manner.

### 4.5.1. Application Programming Interface

An Application Programming Interface (API) is a particular set of rules ('code') and specifications that software programs follow to communicate with each other (figure 9). It serves as an interface between different software programs and facilitates their interaction; similar to the way the consumer interface facilitates interaction between humans and computers. In fact, most commercial monitoring tools such as Rackspace, Nimsoft, RevealCloud, and LogicMonitor provide their consumers with extensible open APIs to enable them specifying their own required system functionalities.

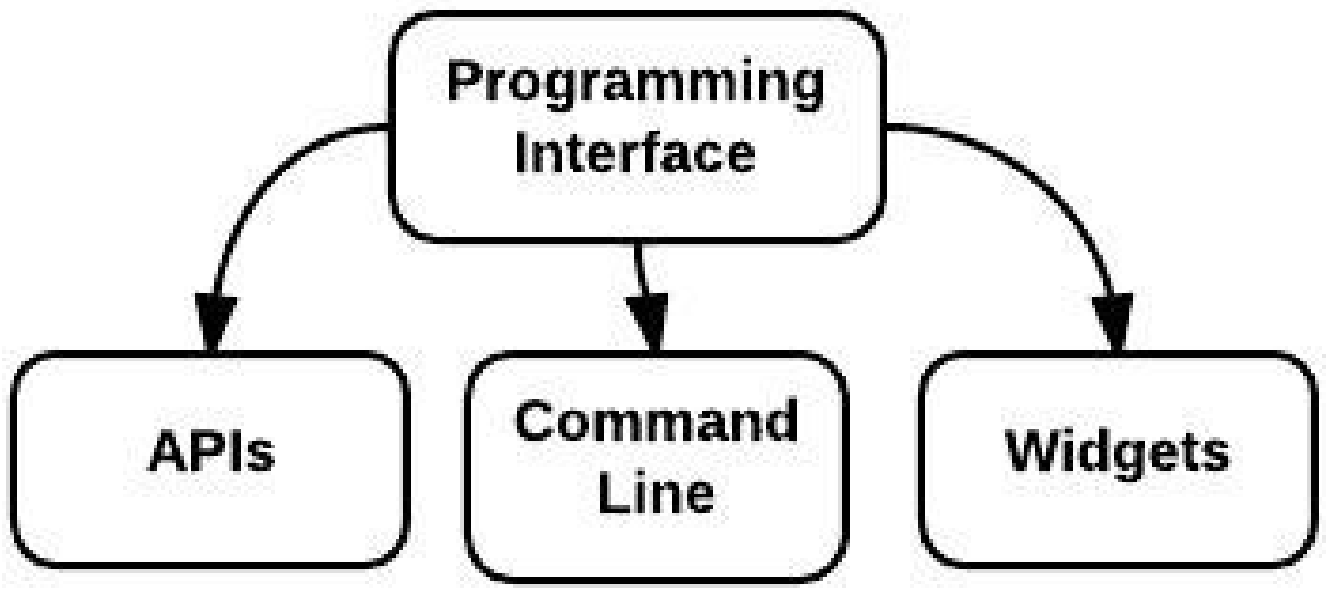

**Fig. 9: Different Types of Programming Interfaces.**

### 4.5.2. Command-line

A command line provides a means of communication between a consumer and a computer that is based solely on textual input and output.

### 4.5.3. Widgets

In computer software, a widget is a software service available to consumers for running and displaying applets on a graphical interface on the desktop. Monitis and RevealCloud are two popular commercial tools that provide a consumer data on multiple customizable widgets.

### 4.5.4. Communication Protocols

All commercial tools adopt communication protocols for data transfer in and out. Communication protocols vary and are different from a monitoring tool to another. For example, Monitis and Rackspace follow HTTPs and FTP protocols. Another example is LogicMonitor, which adopts the encrypted Simple Network Management Protocol (SNMP).

# 5. Commercial Monitoring Tools

## 5.1. Monitis

Monitis [33] founded in 2005, has one unified dashboard where consumers can open multiple widgets for monitoring. A Monitis consumer needs to enter his\her credentials to access the hosting cloud account. In addition, a Monitis consumer can remotely monitor any website for uptime, in-house servers for CPU load, memory, or disk I/O, by installing Monitis agents to retrieve data about the devices. A Monitis agent can also be used to collect data of networked devices in an entire network (behind a firewall), this technique is used instead of installing a Monitis agent on each single device. Widgets can be also emailed as read only version to share the monitored information. Moreover, Monitis provides rich features for reporting the status of instances where consumers can specify the way a report should be viewed e.g. chart, or graph. It also enables its consumers to share the report publicly with others.

## 5.2. RevealCloud

CopperEgg [36][37] provides RevealCloud monitoring tool. It was founded in 2010 and Rackspace is a main partner. RevealCloud enables its consumers to monitor across cloud layers e.g. SaaS, PaaS, and IaaS. It is not dedicated to only one cloud resources provider, rather it is generic to allow a consumer to get its benefits within most popular cloud providers e.g. AWS EC2, Rackspace, etc. RevealCloud is one of the very few monitoring tools that supports maintaining monitored historical data, it can trace up to last 30 days data, which is considered as a prime feature that most commercial monitoring tools lack.

## 5.3. LogicMonitor

LogicMonitor [35] was founded in 2008 and it is a partner with several third parties such as NetApp, VMWare, Dell, and HP. Similarly to RevealCloud, LogicMonitor enables its consumers to monitor across cloud layers e.g. SaaS, PaaS, and IaaS. It also enables them to operate monitoring operations on multi-cloud resources. Protocol used in communications is SSL outgoing only encrypted connections. Moreover, LogicMonitor uses Simple Network Management Protocol (SNMP) as a method of retrieving data about distributed virtual and physical resources.

## 5.4. Nimsoft

Nimsoft was founded in 2011 [34]. Nimsoft supports multi-layers monitoring and both virtual and physical cloud resources. Moreover, Nimsoft enables its consumers to view and monitor their resources in case they are hosted on different cloud infrastructures e.g. a Nimsoft consumer can view resources on Google Apps, Rackspace, Amazon, Salesforce.com and others through a unified monitoring dashboard. Also, Nimsoft gives its consumers the ability to monitor on both private and public clouds.

## 5.5. Nagios

Nagios was founded in 2007 [38], it supports multi-layer monitoring. It enables its consumers to monitor their resources on different cloud infrastructures as well as in-house infrastructure. Nagios utilizes SNMP for monitoring networked resources. Moreover, Nagios has been extended with monitoring functionalities for both virtual instances and storage services using a plugin-based architecture [20]. Typically, a Nagios server is required to collect the monitoring data, which would place it as a centralized solution. Moreover, Nagios is a cloud solution as a user would need to setup a Nagios server. However, many possible configurations can help create multiple hierarchical Nagios servers to reduce the disadvantages of a centralized server.

## 5.6. SPAE by SHALB

SHALB was founded in 2002 [39] and provides a monitoring solution called SPAE (Security Performance Availability Engine). SPAE is a typical network monitoring tool supporting a variety of network protocols such as HTTP, HTTPS, FTP, SSH, etc. It uses SNMP [40] to perform all of its monitoring processes and emphasizes security monitoring and vulnerability. However, SPAE does not support monitoring at different layers (IaaS, PaaS and SaaS). It enables its consumers to monitor networked resources including cloud infrastructure.

## 5.7. CloudWatch

CloudWatch [41] is one of the most popular commercial tools for monitoring the cloud. It is provided by Amazon to enable its consumers monitoring their resources residing on EC2. Hence, it does not support multi-cloud infrastructure monitoring. The technical approaches used in CloudWatch to collect data are implicit and not exposed to users. CloudWatch is limited in monitoring resources across cloud layers. However, an API is provided for users to collect metrics at any cloud layer but requires the users to write additional code.

## 5.8. OpenNebula

OpenNebula [42] is an open source monitoring system that provides management for data centers. It uses SSH as the protocol permitting consumers to gain access and gather information about their resources. Mainly, OpenNebula is concerned with monitoring physical infrastructures involved in data centers such as private clouds.

## 5.9. CloudHarmony

CloudHarmony [43] started monitoring services in the beginning of 2010. It provides a set of performance benchmarks of public clouds. It is mostly concerned in monitoring the common operating system metrics that are related to (CPU, disk and memory). Moreover, cloud to cloud network performance in CloudHarmony is evaluated in terms of RTT and throughput.

## 5.10. Windows Azure FC

Azure Fabric Controller (Azure FC) [44][45] is adopting centralized network architecture. It is a multi-layer monitoring system but, it does not support monitoring across different cloud infrastructures. Moreover, Azure FC utilizes SNMP for performing monitoring.

# 6. Conclusion, Discussion and Future Research Directions

This paper presented and discussed the state-of-the-art research in the area of cloud monitoring. In doing so, it presented several design issues and research dimensions that could be considered to evaluate a cloud computing system. It also presented several cloud monitoring tools, their features and shortcomings. Finally, this paper presented future research directions that should be considered to develop efficient cloud monitoring systems.

With increasing cloud complexity, efforts needed for management and monitoring of cloud infrastructures need to be multiplied. The size and scalability of clouds when compared to traditional infrastructure involves more complex monitoring systems that have to be more scalable, effective and fast. Technically, this would mean that there is a demand for real-time reporting of performance measurements while monitoring cloud resources and applications. Therefore, cloud monitoring systems need to be advanced and customized to the diversity, scalability, and high dynamic cloud environments.

In section 4, we analyzed in detail the main evaluation dimensions of monitoring. As discussed, not all of those dimensions are adopted by monitoring systems in either open source or commercial domains. Though, most of these dimensions, which are basically related to performance, have been addressed by the research community and have received attention. However, more considerable effort to achieve the maturity level is essential for monitoring cloud systems.

Decentralized approaches are gaining more trust over centralized approaches. In contrast to unstructured P2P, structured P2P networks present a practical and more efficient approach in terms of network architecture. However, considerable study is needed on decentralized networks that are with various degrees of centralization. Considering interoperability, either cloud-dependent or cloud-agnostic, both of these monitoring approaches gain high importance. Currently, both approaches are supported by several monitoring systems. Through our study, we found that cloud-dependent monitoring systems are mostly commercial, whereas, cloud-agnostic monitoring systems are typically open source.

We observe that matrix of the quality of service is the most important dimension of a monitoring system and list the quality parameters that can be monitored along with the related criteria. We also elaborate on how those quality parameters should be monitored, detected and reported. At which cloud layer a monitoring system should operate the monitoring processes. Further, the aggregation of multiple parameters for a consumer application is a critical aspect of monitoring. This means that a monitoring system be a cloud layer specific or layer agnostic. This will determine the visibility characteristic of a cloud monitoring system. All of these issues in monitoring need more study by the cloud community and still in demand of more technical improvements. Table 4 summarizes our study of monitoring platforms against evaluation dimensions explored in section 4.

**Table 4: Monitoring Platforms against evaluation Dimensions**.

| **Platform** | Network Arch. (Centralized) | Network Arch. (Decentralized) | Interoperability Multi-Cloud | Visibility Multi-Layers | SNMP | Extendable APIs |
|---|---|---|---|---|---|---|
| Monitis [33] | Not-Stated (SaaS solution) | Not-Stated (SaaS solution) | Yes | Yes | Yes | Yes |
| RevealCloud [36][37] | Not-Stated (SaaS solution) | Not-Stated (SaaS solution) | Yes | Yes | Not-Stated | Yes |
| LogicMonitor [35] | Not-Stated (SaaS solution) | Not-Stated (SaaS solution) | Yes | Yes | Yes | Yes |
| Nimsoft [34] | Yes | Yes | Yes | Yes | Yes | Yes |
| Nagios [20][38] | Yes | Yes | Yes | Yes | Yes | Yes |
| SPAE [39][40] | Not-Stated (SaaS solution) | Not-Stated (SaaS solution) | Yes | No | Yes | No |
| CloudWatch [41] | Not-Stated (SaaS solution) | Not-Stated (SaaS solution) | No | Yes | Not-Stated | Yes |
| OpenNebula [42] | Yes | No | No | No | Not-Stated | No |
| CloudHarmony [43] | Not-Stated (SaaS solution) | Not-Stated (SaaS solution) | Yes | No | Not-Stated | No |
| Azure FC [44][45] | Yes | Not-Stated | No | Yes | Yes | Yes |

Since monitoring becomes an essential component of the whole cloud infrastructure, its elasticity has to be given a high considerable priority. Based on this fact and on the aforementioned monitoring aspects and approaches, we believe that considerable effort is required to have more reliable cloud monitoring systems. Furthermore, we found there is a lack of reachable standards on procedure, format, and metrics to assess the development of cloud monitoring. Hence, we recommend having more collaborative use of research facilities in which tools, lessons learned and best practices can be shared among all interested researches and professions.

## 7. Conclusion